\documentclass[conference,a4paper]{IEEEtran}
\IEEEoverridecommandlockouts
\usepackage{cite}
\usepackage{amsmath,amssymb,amsfonts}
\usepackage{algorithmic}
\usepackage{graphicx}
\usepackage{textcomp}
\usepackage{xcolor}
\usepackage{hyperref}
\usepackage{url}
\usepackage{soul} 

\def\BibTeX{{\rm B\kern-.05em{\sc i\kern-.025em b}\kern-.08em
    T\kern-.1667em\lower.7ex\hbox{E}\kern-.125emX}}
\begin{document}

\title{Validation, Verification, and Testing (VVT) of future RISC-V powered cloud infrastructures: the Vitamin-V Horizon Europe Project perspective \thanks{Funded by the European Union. Views and opinions expressed are, however, those of the authors only and do not necessarily reflect those of the European Union or the HaDEA. Neither the European Union nor the granting authority can be held responsible for them. Project number: 101093062}}

\author{\IEEEauthorblockN{Martí	Alonso\textsuperscript{1}, David Andreu\textsuperscript{1}, Ramon Canal\textsuperscript{1}, Stefano Di Carlo\textsuperscript{2}, Cristiano Chenet\textsuperscript{2}, Juanjo Costa\textsuperscript{1},\\ Andreu Girones\textsuperscript{1}, Dimitris Gizopoulos\textsuperscript{3}, Vasileios Karakostas\textsuperscript{3}, Beatriz	Otero \textsuperscript{1}, 
George Papadimitriou\textsuperscript{3},\\ Eva Rodríguez\textsuperscript{1},  Alessandro Savino\textsuperscript{2}}
\IEEEauthorblockA{\\
\textsuperscript{\textbf{1}}Universitat Politècnica de Catalunya, Barcelona, Spain\\
\textsuperscript{\textbf{2}}Politecnico di Torino, Torino, Italy \\
\textsuperscript{\textbf{3}}University of Athens, Athens, Greece \\ 
Contact email: stefano.dicarlo@polito.it
}}

\maketitle

\begin{abstract}
Vitamin-V is a project funded under the Horizon Europe program for the period 2023-2025. The project aims to create a complete open-source software stack for RISC-V that can be used for cloud services. This software stack is intended to have the same level of performance as the x86 architecture, which is currently dominant in the cloud computing industry. In addition, the project aims to create a powerful virtual execution environment that can be used for software development, validation, verification, and testing. The virtual environment will consider the relevant RISC-V ISA extensions required for cloud deployment.
Commercial cloud systems use hardware features currently unavailable in RISC-V virtual environments, including virtualization, cryptography, and vectorization. To address this, Vitamin-V will support these features in three virtual environments: QEMU, gem5, and cloud-FPGA prototype platforms. The project will focus on providing support for EPI-based RISC-V designs for both the main CPUs and cloud-important accelerators, such as memory compression. The project will add the compiler (LLVM-based) and toolchain support for the ISA extensions.
Moreover, Vitamin-V will develop novel approaches for validating, verifying, and testing software trustworthiness. This paper focuses on the plans and visions that the Vitamin-V project has to support validation, verification, and testing for cloud applications, particularly emphasizing the hardware support that will be provided.
\end{abstract}



\bstctlcite{IEEEexample:BSTcontrol}

\section{Introduction}

RISC-V is a revolutionary open-source instruction set architecture (ISA) designed to offer simplicity, modularity, and extensibility \cite{Waterman:EECS-2016-129}. This exciting development brings many benefits over proprietary processor architectures, including the potential for customization and lower licensing costs \cite{greengard2020will}.

Despite these advantages and the fact that RISC-V applications have started to see their birth in the embedded domain \cite{10.1145/3457388.3458657}, several challenges still need to be addressed before RISC-V can be widely adopted for cloud applications.
One key obstacle is the maturity of the RISC-V ecosystem. The platform has gained significant momentum in recent years, but the ecosystem surrounding RISC-V processors is still developing. This includes hardware and software tools and the number of vendors and support services available \cite{9771410}. As the ecosystem continues to mature, it is expected that this will become less of a concern.

Another potential challenge is performance. Although RISC-V processors can offer good performance, they may not yet be able to match the performance of more established architectures, such as x86 or ARM, in specific applications. This could limit the adoption of RISC-V in performance-sensitive cloud applications. As technology continues to evolve, this may become less of a barrier.

Compatibility is another potential challenge to widespread RISC-V adoption. Many cloud applications are designed to run on x86 or ARM architectures and may not be compatible with RISC-V processors. This could limit the use of RISC-V in specific cloud environments. Efforts are underway to address this issue by developing emulation and virtualization solutions.

Security is also a concern. As RISC-V processors become more widely adopted, there is an increasing potential for security attacks. Ensuring the security of RISC-V-based cloud applications will be an important challenge that needs to be addressed as technology develops.

Eventually, standardization is another area that needs to be addressed. While RISC-V is an open standard, there is still a need for further standardization in areas such as memory management and I/O interfaces. This can lead to compatibility issues between different RISC-V implementations and limit the portability of RISC-V-based cloud applications. Efforts are underway to address this issue by developing standardization solutions that can promote interoperability.

Vitamin-V is a research project funded by the European Commission in the 2023-2025 time frame to propose innovative solutions that aim to address these challenges. Vitamin-V will deploy a complete RISC-V hardware-software stack for cloud services based on cutting-edge cloud open-source technologies for RISC-V cores, with a focus on EPI cores. Vitamin-V is designed to offer an innovative RISC-V virtual execution environment, which provides hardware emulation, simulation, and FPGA prototyping to enable software development, verification, and validation before actual hardware is released. Additionally, Vitamin-V contributes to porting the complete cross-compiling toolchain, software stack, and essential application libraries for the forthcoming release of the RISC-V EPI processors \cite{EPI}. This solution is expected to play a critical role in promoting the adoption of RISC-V for cloud applications \cite{10.1145/3310273.3323432}.

In particular, Validation, Verification, and Testing (VVT) activities are among the most critical activities during software development and deployment, with potential risks in terms of the safety and security of cloud applications. After providing a general overview of the Vitamin-V activities, this paper wants to focus on plans and visions that the Vitamin-V project has to support VVT activities for cloud applications, concentrating on the support that the hardware itself can provide to these activities.

The paper is organized as follows: Section \ref{sec:concept} overviews the Vitamin-V project organization, while Section \ref{sec:VVT} provides a deeper overview of the project's implementation plans VVT techniques for RISC-V cloud applications. Finally, Section \ref{sec:conclusions} summarizes the main contributions of the paper.

\section{Concept and Methodologies}
\label{sec:concept}
\begin{figure}[hbt]
\centering
\includegraphics[width=8cm]{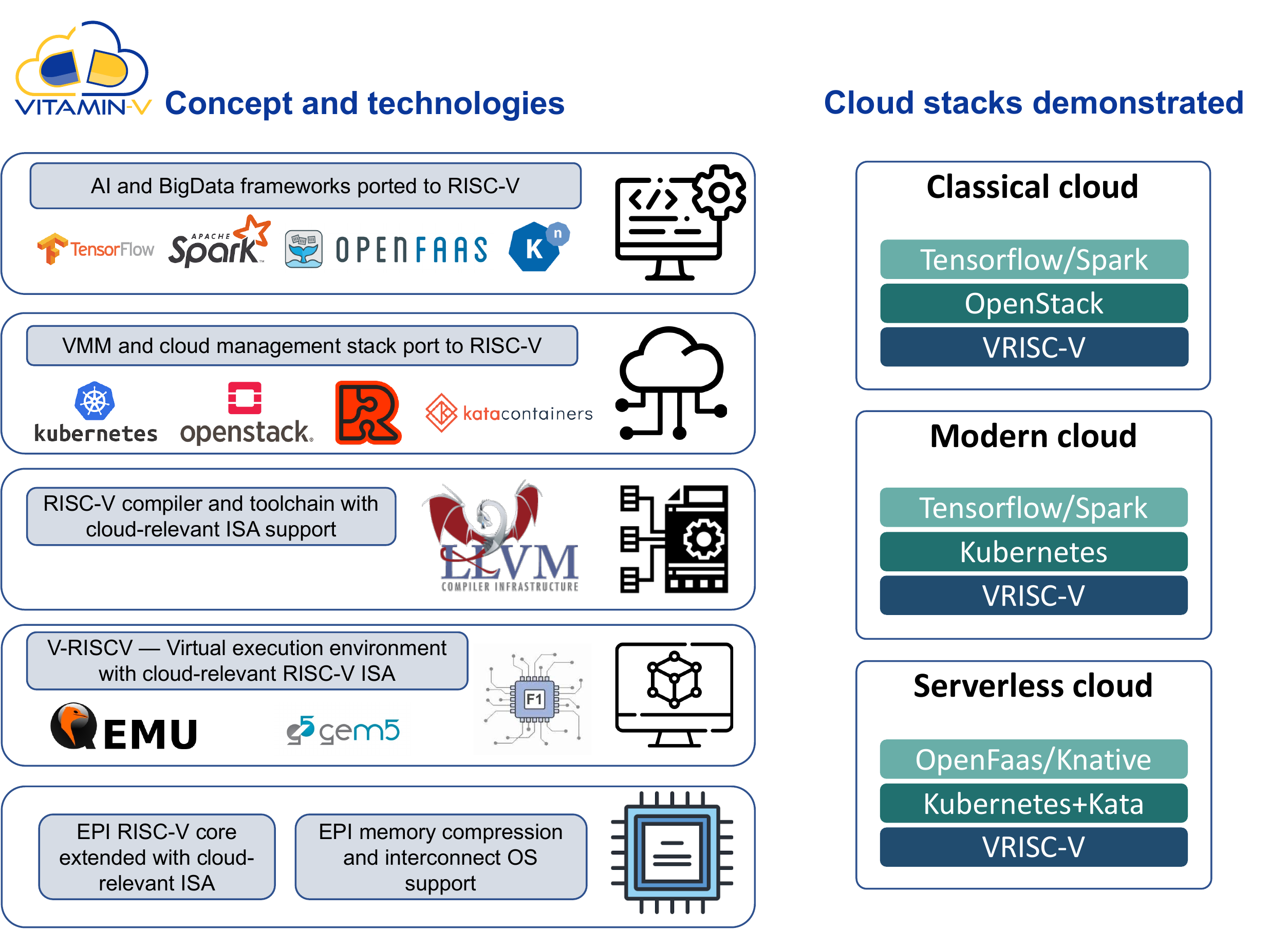}
  \caption{VITAMIN-V concept and architecture}
  \label{fig:concept}
\end{figure}

The objective of Vitamin-V is to create a complete RISC-V cloud software stack that can compete with the dominant x86 counterpart in terms of performance, as shown in Figure \ref{fig:concept}. However, the lack of full-fledged RISC-V systems presents a significant challenge for porting and evaluating advanced cloud setups and software stacks.  Commercial cloud systems use hardware features partially available in RISC-V virtual environments and commercial hardware cores. These features include virtualization, cryptography, and vector extensions.

To address this challenge, Vitamin-V aims to develop the \emph{Vitamin RISC-V virtual execution environment} (VRISC-V), which is a multi-layered high-performance RISC-V virtual environment based on three cutting-edge technologies: functional emulation (QEMU \cite{QEMU}), cycle-accurate simulation (gem5) \cite{gem5}, and an FPGA-based hardware prototype system node capable of running on AWS EC2 F1 FPGAs (and thus scalable to 100s of nodes) (FPGA \cite{Amazon}). These technologies provide unique features essential for software development, validation, verification, and testing, making it easier for software developers to adopt RISC-V. With QEMU and the RISC-V on FPGA, the project aims at simulating multi-node systems with more than 100 cores. The goal is to simulate multi-node systems in gem5 with cycle-accurate accuracy and microarchitectural configuration flexibility \cite{8491804}. 

Vitamin-V will support accelerators, such as memory compression, in the VRISC-V virtual execution environment and provide a mature compiler toolchain based on LLVM \cite{LLVM:CGO04} to handle the complete RISC-V ISA, including its extensions. Additionally, the project will develop a validation, verification, and testing (VVT) toolset to identify software bugs and malicious code sequences to ensure the trustworthiness of the software layers in future RISC-V systems. The VVT tools will use the VRISC-V platform to facilitate software porting and prototyping before mature hardware is available. This will promote the migration towards new RISC-V servers in the cloud.

To enable the execution of complete cloud stacks on the VRISC-V virtual execution environment, Vitamin-V will port all necessary machine-dependent modules in relevant open-source cloud software distributions. These modules include support for running entire Virtual Machines (VMs), containers, and lightweight VMs (KVM, QEMU, Docker, RustVMM), safety-security trusted execution environments (VOSySMonitoRV \cite{caforio2021vosysmonitorv}), cloud management software (OpenStack, Kubernetes, Kata Containers), and AI and Big Data libraries (Tensorflow, Spark). 

The project will address classic cloud stacks that target the execution of entire VMs managed by OpenStack, modern cloud setups that target entire VMs and containers managed by Kubernetes, and serverless cloud stacks that target the execution of lightweight VMs managed by Kubernetes with Kata Containers. Vitamin-V will benchmark the three working cloud setups against relevant AI applications (i.e., Google Net, ResNet, VGG19), Big-Data applications (TPC-DS on top of Apache Spark), and Serverless applications (FunctionBench, ServerlessBench). The goal is to match the software performance of its x86 equivalent, using CPU core mark scores for a fair assessment. Overall, this development will establish a RISC-V cloud-stack ecosystem for market adoption.

\section{Validation, Verification, and Testing for RISC-V cloud services}
\label{sec:VVT}

In the age of cloud services, it is often difficult to transfer software workloads between different computing platforms without sacrificing performance and trustworthiness \cite{tampouratzis2018novel}. Several factors can contribute to decreased performance when moving from one architecture to another, including poor software implementation, inefficient data structures, and limited use of caching. Additionally, the software can be compromised by bugs or malicious code at any point in its lifespan. As a result, to optimize and ensure the trustworthiness of an application, it is essential to monitor its execution, identify performance bottlenecks, and customize the software to fit the underlying hardware. 
However, this task cannot be accomplished using simple performance metrics such as execution time or clock cycles. Multiple factors come into play when mapping software to a modern computing system, including in- vs. out-of-order execution engines, pipeline stages, execution ports and corresponding latencies, re-order buffers, load/store queues, and cache organization. As a result, capturing and analyzing detailed performance metrics is becoming increasingly necessary to enable in-depth architecture modeling and optimization procedures \cite{marques2020application}.

This section explains how the Vitamin-V project intends to utilize the RISC-V Hardware Performance Monitor (HPM) unit to implement advanced Validation, Verification, and Testing for RISC-V cloud services.

\subsection{Hardware Performance Monitoring}

Like other modern processors, RISC-V processors come with hardware performance monitoring units to keep track of processor performance. These units have become necessary due to the increased complexity of processors in recent decades, which has resulted in the need for hierarchical cache subsystems, non-uniform memory, simultaneous multithreading, and out-of-order execution. Software that can understand and adjust to resource utilization has benefits for performance and efficiency.

The RISC-V ISA has a simpler HPM unit than x86, but it defines a flexible and open-source performance monitoring solution that can be implemented in various ways. Vitamin-V plans to support the latest RISC-V HPM specifications, starting with hardware simulation in VRISC-V and integrating it into the operating system using open-source libraries like Linux perf \cite{rperf} and PAPI \cite{1592755}. The implemented HPM will have hardware registers and counters for microarchitectural events that software can access and hardware assertions from gem5 useful for debugging~\cite{lowepower2020gem5}. Vitamin-V aims to build at least essential counter events during early development, using existing RISC-V cores like Semidynamics’ Atrevido Core~\cite{atrevido}.

The hardware performance monitoring units track various events related to the processor's architecture and micro-architecture, such as retired instructions, branch predictions, cache hits and misses, floating-point operations, hardware interrupts, elapsed core clock ticks, and core frequency. These events generate several parameters, but the processors have only a few registers to store them, so only a few hardware performance counters are available at any given time. The design complexity and cost of concurrent monitoring of events limit the number of available hardware performance counters \cite{Sprunt_2002,Malone_2011,Doyle_2017}.

Performance monitoring instructions allow developers to access the counters by reading or writing their values. Vitamin-V will investigate VRISC-V's hardware monitor extensions to address the need for software validation, verification, and testing in cloud environments. The project plans to use hardware monitoring to assess the trustworthiness of software deployed on RISC-V. Vitamin-V will implement various monitoring events to support software validation, verification, and testing for future implementation in RISC-V cores, using the state-of-the-art gem5 simulator \cite{gem5}.

\subsection{VVT through Static Analysis}

It is crucial to evaluate the trustworthiness of software at an early stage before it is executed. Vitamin-V is planning to implement a machine learning~(ML) tool to analyze the static content of executable files to determine whether they are benign or malicious. Previous work on using static analysis of executable files to detect software bugs and security threats \cite{ding2014control, haddadpajouh2018deep} has inspired Vitamin-V's decision to incorporate deep learning~(DL) techniques to identify complex patterns from a vast amount of labeled data. However, there are situations where the dataset may not be sufficient or a zero-day attack detection is required. In such cases, a transfer
learning~(TL) technique can be implemented to enhance the accuracy of the detection process. Transfer learning involves reusing a pre-trained model for a different task~\cite{Rodriguez2022}.

\begin{figure}[bt]
    \centering
  \includegraphics[width=.75\linewidth]{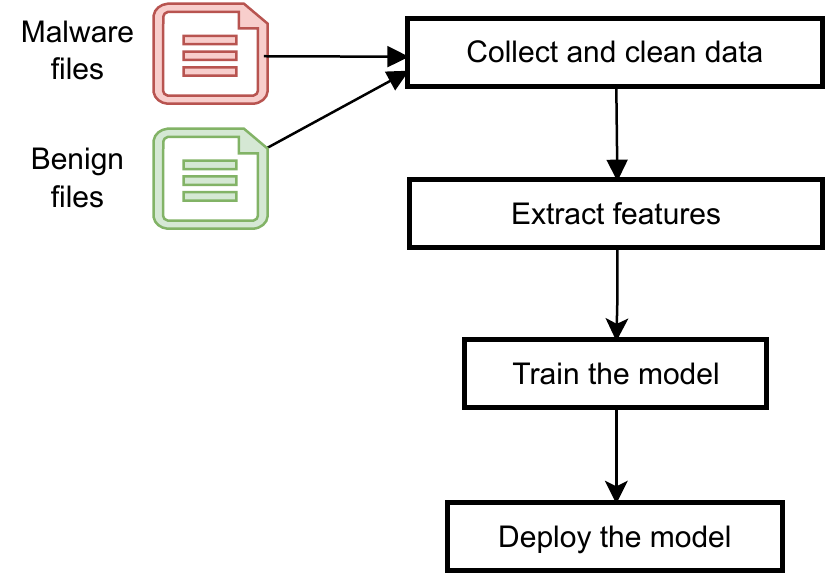}
    \caption{Methodology used to create the Deep Learning model used in VITAMIN-V.}
  \label{fig:staticPhases}
\end{figure}

The proposed methodology is based on a similar approach to the one used by Haddadpajouh et al. \cite{haddadpajouh2018deep} and consists of four main steps as illustrated in Figure~\ref{fig:staticPhases}: dataset creation, feature extraction, model training, and model deployment and tuning.

The first step in training any neural network model is to have a balanced collection of benign and malicious codes. Vitamin-V focuses on detecting various types of malicious codes, including hardware attacks like spectre \cite{Kocher2019}, meltdown \cite{Lipp2018}, viruses, and other malware. To achieve this, Vitamin-V will gather different source codes from popular public repositories such as VirusTotal \cite{VirusTotal}, VirusShare \cite{VirusShare}, or SourceFinder \cite{rokon2020sourcefinder}. However, since there is currently a lack of source code repositories for RISC-V malicious codes, the plan is to use the AMD64 version of the collected malware and recompile them for the target platform. This approach is suitable for the current situation, while new malicious codes for the RISC-V architecture can be added to the dataset when they become available. Vitamin-V plans to download a selection of common Linux application packages from the Debian repository, which already has RISC-V versions, for the benign codes. A balanced selection of both sets of codes will be built.

Each program must be represented as a feature vector to classify executables. Vitamin-V uses Linux as its operating system; therefore, the binary files use the Executable and Linking Format (ELF). These ELF binaries can be disassembled using default Linux tools to a textual format and extract their sequence of instructions (the operation codes or OpCodes). This sequence of instructions can be analyzed to obtain the frequency of different OpCodes sequences (the n-grams) and then calculate the feature vector for each sequence, which will help train the neural network.

The categorized feature vectors can be fed to the DL network and train it to generate a model capable of detecting benign or malicious codes. However, due to the previously mentioned lack of malicious codes in RISC-V, the resulting model will -initially- only be able to detect malicious code coming from AMD64 platforms. 

To overcome this limitation, Vitamin-V plans to use transfer learning techniques to transfer this knowledge to a new neural network that will detect malicious code and zero-day attacks on RISC-V binaries. To fine-tune the model, new data will be collected, such as hardware performance counters, to adjust the parameters of the pre-trained model and improve its performance in detecting malware attacks. After an evaluation period to ensure the proper performance of the detection mechanism, the proposed model will be deployed. Additional data collection and fine-tuning steps may be required to keep the model accurate and effective over time.

\subsection{Dynamic Analysis using Hardware Performance Counters}

\begin{figure}[hbt]
    \centering
    \includegraphics[width=0.8\columnwidth]{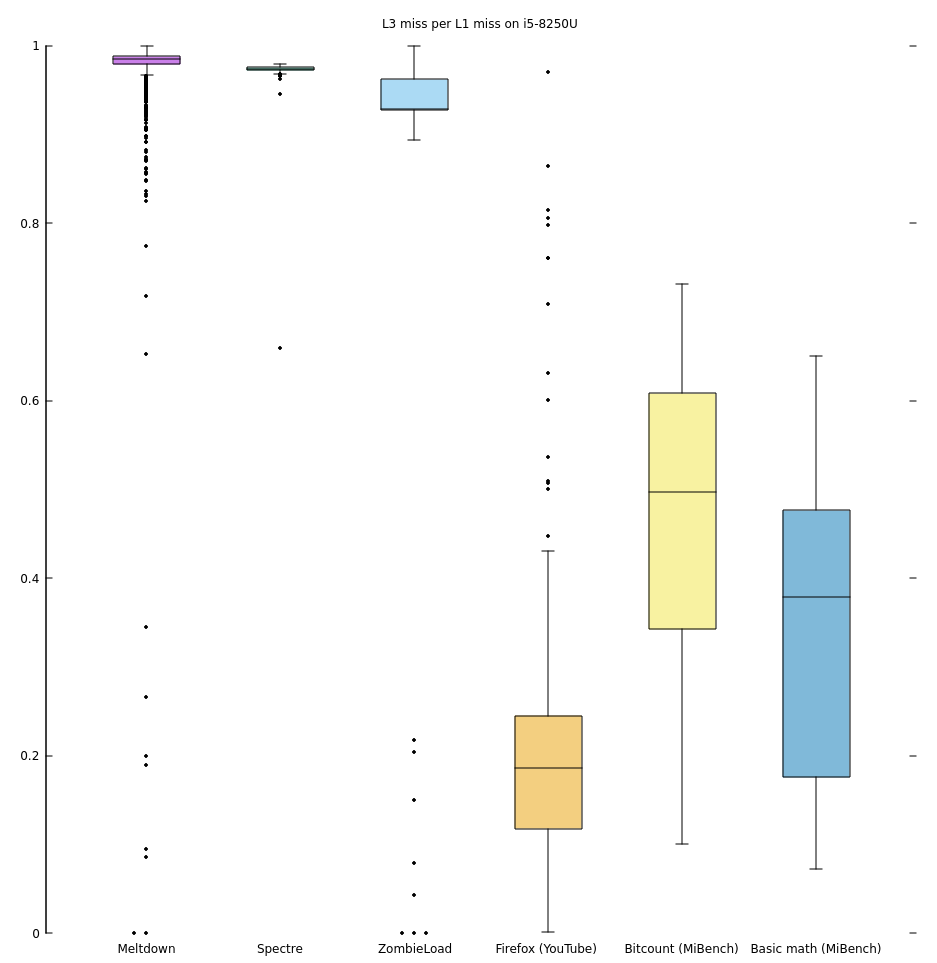}
    \caption{Using hardware counters to detect benign and malicious code.}
  \label{fig:useHWCounters}
\end{figure}

Although an application may be deemed trustworthy through static verification, corruption can still occur during runtime, necessitating dynamic monitoring. Such corruption can result from environmental conditions, electromagnetic fields, space radiation, aging, design flaws, manufacturing imperfections, and intentional attacks, which can generate abnormal behavior in processors, compromising the safety, reliability, and security of modern computing systems. Anomaly detection involves identifying patterns in data that deviate from expected behavior \cite{Chandola_2009}. Vitamin-V intends to exploit the RISC-V HPM unit for this purpose, as demonstrated by the proof of concept reported in Figure~\ref{fig:useHWCounters}, where the use of hardware counters distinguishes between benign and malicious applications. The figure shows a box plot of the frequency of a specific ratio, L3 misses per L1 misses, clearly identifying three malicious codes (MeltDown, Spectre, and ZombieLoad) in contrast to three benign applications (two applications from the MiBench benchmark and a YouTube video visualization in a Firefox browser).

Anomaly detection through hardware and artificial intelligence (AI) involves the dynamic analysis of micro-architectural events in a processor by machine learning (ML) algorithms, which distinguish between normal and anomalous behavior. This approach was first introduced in the hardware-based malware detector proposed by Demme et al. in 2013 \cite{Demme_2013} and exploited for soft error detection in \cite{9474120,kasap2023microarchitectural}.

The intuition behind detecting anomalies based on hardware performance counters (HPCs) stems from the fact that programs exhibit phase behavior \cite{Sherwood_2003,Isci_2006}. Programs perform activities in distinct phases, which can correspond to patterns in architectural and micro-architectural events, thus enabling the detection of anomalies based on time-behavioral patterns of the HPCs.

\begin{figure}[htb]
\centerline{\includegraphics[width=\columnwidth]{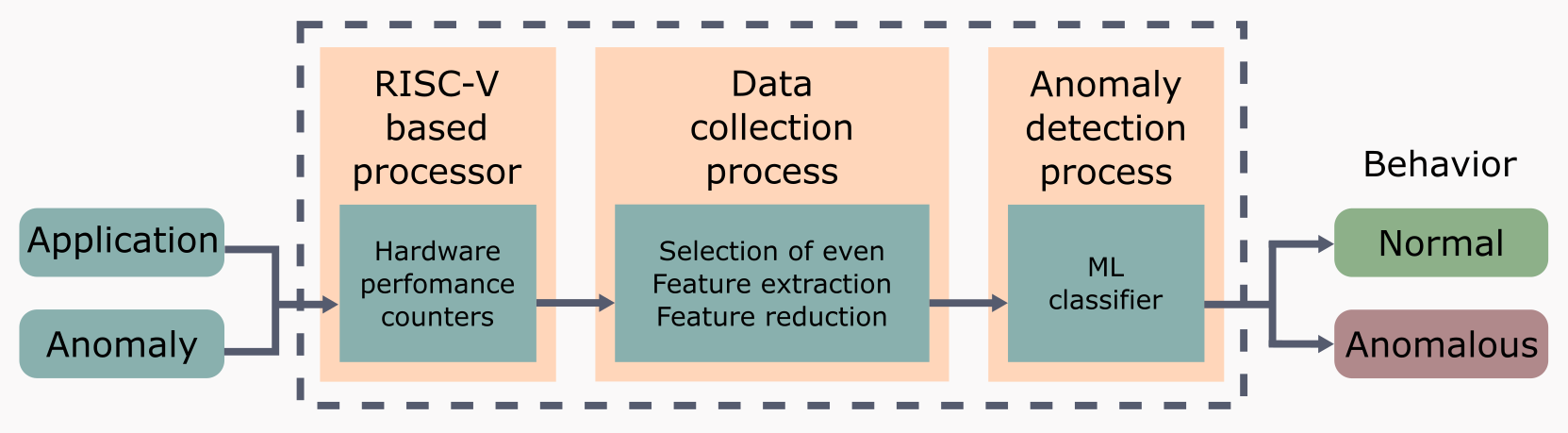}}
\caption{The hardware and AI-based anomaly detector framework. Elaborated by the author.}
\label{fig:Anomaly_detector_framework}
\end{figure}

The Vitamin-V hardware and AI-based anomaly detector framework is presented in \autoref{fig:Anomaly_detector_framework}, with its primary goal of differentiating between normal execution and anomalous execution. The framework comprises three fundamental blocks: (i) the RISC-V based processor with its HPCs, (ii) the data collection process, and (iii) the anomaly detection itself, performed by an ML classifier. The detector's performance and efficiency are evaluated to determine its effectiveness.

Adding hardware performance counters presents challenges due to the design complexity and cost of monitoring events under speculative execution. Mobile and Internet of Things devices have even stricter resource constraints, making adding more hardware performance counters challenging. The trade-off between the number of hardware events and detection accuracy is important to consider. To achieve reasonable accuracy, some works analyze more events than the number of available hardware performance counters, requiring running the application multiple times \cite{Demme_2013,Singh_2017,Sayadi_2017}. This limits the run-time applicability of hardware-based anomaly detection. A careful design of the machine learning classifier is needed to compensate for the reduced characterization due to the fewer events analyzed.

The process of collecting data involves selecting events, extracting features, and reducing them \cite{Chenet_2023}. In the hardware and AI-based anomaly detection context, selecting events consists of choosing from several hardware events available in the processor for use in the anomaly detector. Feature extraction is the process of capturing and storing the registered hardware performance counters (HPCs) in a vector space for analysis by the machine learning (ML) model. Feature reduction, also known as dimensionality reduction, is a form of data processing that deals with redundant dimensions in high-dimensional space. Redundant dimensions contribute to the measurement of noise in the training dataset, which reduces the detection rates of testing.

The selection of events and feature reduction aims to select the most relevant/predictive data for anomaly detection. Since only a few hundred hardware events are available in the processors, it is possible to perform manual event selection through empirical knowledge of each event's representation in the architecture and micro-architecture or based on other studies. However, in some cases, researchers may need this information and start collecting all the hardware events available in the processor, generating big data. In such cases, manual analysis is impossible, and some feature selection techniques must be used. Some commonly used methods include Principal Component Analysis (PCA), Fisher Score \cite{Duda_2000}, Pearson Correlation Coefficient \cite{Pearson_1895} based-techniques, and Information Gain (also called Mutual Information) \cite{Peng_2005}.

A key characteristic of feature extraction is the sampling period of the HPC. There is no set rule for this value. Still, in hardware-based detection experiments, they generally remain in the order of milliseconds or seconds or even as multiples of processor cycles or instruction epochs.

The third fundamental block in the framework is anomaly detection itself, performed by ML classifiers. After training with input data, these classifiers can automatically categorize data into one or more classes (since the training and tested data have similar statistical distributions). Classifiers may differ in terms of the labels available (multi-class and one-class), the way they learn (supervised, unsupervised, and semi-supervised), the mathematical formula/algorithms they implement (Neural Networks-Based, Linear Models-Based, Decision Trees-based, Rule-Based, etc.), and the arrangement of classifiers (multiple stages). As each classifier configuration delivers different results across various metrics (including performance, efficiency, and hardware design overhead), designing an ML classifier is critical.

Multi-class classification-based anomaly detection assumes that the data includes instances labeled as belonging to multiple normal classes during training. When a new instance is tested, it is considered anomalous if it is not classified as normal by any of the trained classifiers. On the other hand, one-class classification-based anomaly detection techniques assume that all the training instances have only one class label, and any new instance that falls outside of the learned boundary is declared anomalous \cite{Chandola_2009}.

Supervised learning refers to classifiers trained using labeled examples, while unsupervised learning uses unlabeled examples. Semi-supervised learning combines labeled and unlabeled examples in the dataset \cite{Burkov_2019}. However, obtaining labeled data can be expensive, especially when considering abnormal behaviors, which are often dynamic and challenging to label accurately. Therefore, unsupervised learning classifiers are more commonly used for anomaly detection \cite{Chandola_2009}.

Eventually, to improve the accuracy of the classifiers, more complex algorithms such as Ensemble Learning techniques are used \cite{Sayadi_2018}. Ensemble Learning is a branch of machine learning that combines the results of multiple base learners to improve decision accuracy. Examples of Ensemble Learning algorithms include Boosting (AdaBoost implementation \cite{Freund_Schapire_1997}) and Bagging (Bootstrap Aggregation \cite{Breiman_1996}). Multiple-stage classifiers can also be used to improve the accuracy of anomaly detection \cite{Sayadi_2019}.

\section{Conclusions} 
\label{sec:conclusions}

This paper provided an overview of Vitamin-V, a project funded under the Horizon Europe program from 2023 to 2025. The project aims to create a complete open-source virtualization software stack for RISC-V that can be used for cloud services. Among the different activities, one of the main goals of the project is to create a robust virtual execution environment that can be used for software development, validation, verification, and testing thus increasing the trustworthiness of RISC-V cloud applications. Interested readers may follow the project's latest achievements at \url{https://www.vitamin-v.eu}.

\bibliographystyle{IEEEtran}
\bibliography{bibliography}
\end{document}